\documentclass[pre,twocolumn,amsmath,amssymb,nofootinbib,floatfix,superscriptaddress]{revtex4}

\usepackage{enumerate}
\usepackage{graphicx,bm}
\usepackage[normalem]{ulem}

\usepackage{graphicx,bm,xcolor}
\usepackage{enumerate}
\usepackage{graphicx,bm}
\usepackage[normalem]{ulem}

\begin{document}

\title{Abelian Higgs gauge theories with multicomponent scalar fields\\
  and multiparameter scalar potentials}

\author{Claudio Bonati} 
\affiliation{Dipartimento di Fisica dell'Universit\`a di Pisa
        and INFN Largo Pontecorvo 3, I-56127 Pisa, Italy}

\author{Andrea Pelissetto}
\affiliation{Dipartimento di Fisica dell'Universit\`a di Roma Sapienza
        and INFN Sezione di Roma I, I-00185 Roma, Italy}

\author{Ettore Vicari} 
\affiliation{Dipartimento di Fisica dell'Universit\`a di Pisa,
        Largo Pontecorvo 3, I-56127 Pisa, Italy}

\date{\today}

\begin{abstract}
  We consider multicomponent Abelian Higgs (AH) gauge theories with
  multiparameter scalar quartic potentials, which are extensions of
  the $SU(N)$-invariant AH theories, with a smaller global symmetry
  group.  In particular, we consider an AH model with a two-parameter
  scalar potential and $SO(N)$ global symmetry.  We discuss the
  renormalization-group flow of the $SO(N)$-invariant AH field theory,
  and the phase diagram and critical behavior of a corresponding
  three-dimensional (3D) noncompact lattice AH model.  We argue that
  the phase diagram of 3D noncompact $SO(N)$- and $SU(N)$-symmetric
  lattice AH models are qualitatively similar.  In both cases there
  are three phases: the high-temperature Coulomb phase, and the
  low-temperature molecular and Higgs phases that differ for the
  topological properties of the gauge correlations. However, the main
  features of the low-temperature ordered phases, and in particular of
  the Higgs phase, differ significantly in $SO(N)$ and $SU(N)$
  models. In particular, in $SO(N)$ models they depend on the sign of
  the self-interaction parameter $v$ that controls the symmetry
  breaking from $SU(N)$ to $SO(N)$.  As a consequence, also the
  universal features of the transitions related with the spontaneous
  breaking of the global symmetry (those between the high-temperature
  Coulomb phase and the low-temperature molecular and Higgs phases)
  depend on the sign of $v$.
\end{abstract}

\maketitle


\section{Introduction}
\label{intro}

Many emergent collective phenomena in condensed-matter
physics~\cite{Anderson-book,Wen-book} admit an effective description
in terms of Abelian Higgs (AH) gauge theories, in which charged scalar
fields are minimally coupled with an Abelian gauge field.  The phase
structure and the universal features of the transitions in this broad
class of systems have been extensively studied
~\cite{HLM-74,FS-79,Hikami-80,DH-81,FMS-81,DHMNP-81,CC-82,BF-83,
  FM-83,KK-85,KK-86,BN-86,BN-86-b,BN-87,MU-90,RS-90,MS-90,KKS-94,BFLLW-96,
  HT-96,FH-96,IKK-96,KKLP-98,OT-98, CN-99, HS-00, KNS-02,MHS-02,
  SSSNH-02,SSNHS-03,MZ-03,NRR-03, MV-04,SBSVF-04, NSSS-04, SSS-04,
  HW-05, WBJSS-05, CFIS-05, TIM-05, TIM-06, CIS-06, KPST-06,
  Herbut-book, Sandvik-07, WBJS-08, MK-08, JNCW-08, MV-08, KMPST-08,
  CAP-08, KS-08, ODHIM-09, LSK-09, CGTAB-09, CA-10, BDA-10,
  Sandvik-10, Kaul-12, KS-12, BMK-13, HBBS-13, Bartosch-13,
  HSOMLWTK-13, CHDKPS-13, PDA-13, BS-13, NCSOS-15, NSCOS-15, SP-15,
  SGS-16,WNMXS-17, FH-17, PV-19-CP, IZMHS-19, PV-19-AH3d, SN-19,
  Sachdev-19, FH-19, PV-20-largeNCP, SZ-20, PV-20-mfcp, BPV-20-hcAH,
  BPV-21, BPV-21-bgs, BPV-21-ncAH, WB-21, BPV-22-mpf, BPV-22,
  BPV-23b,SZJSM-23,BPV-23c}, paying particular attention to the role
of the topological features of the gauge correlations, to the
interplay of gauge and scalar excitations, and to the role of the
global symmetry whose spontaneous breaking is crucial for the emerging
Higgs phases, see, e.g., Ref.~\cite{Maas-19}. Several lattice AH gauge
models have been investigated, using both compact and noncompact gauge
variables.  They provide examples of topological transitions, which
are driven by extended charged excitations without a local order
parameter, or by a nontrivial interplay between long-range scalar
fluctuations and nonlocal topological gauge modes.  Most studies on
multicomponent systems have focused on AH theories with $N$-component
fields and global $SU(N)$ symmetry, see, e.g.,
Refs.~\cite{HLM-74,IKK-96,MZ-03,FH-17,BPV-21-ncAH,BPV-22,BPV-23c}.  In
this paper we consider field-theoretical and lattice AH models with
more complex scalar self-interactions, which are invariant under a
smaller group, focusing mainly on systems with a reduced $SO(N)$
invariance.  Some extensions of the $SU(N)$-symmetric AH field
theories have been already discussed in Ref.~\cite{MU-90}. We return
to this issue, and extend the analysis to the phase diagram and
critical behaviors of their lattice counterparts, and in particular to
the $SO(N)$-symmetric lattice AH models with noncompact gauge
variables.

In the $SU(N)$-symmetric AH field theory (UAHFT) an
$N$-component complex scalar field ${\bm \phi}({\bm x})$ is  minimally
coupled with a $U(1)$ gauge field $A_\mu({\bm x})$. The Lagrangian
reads
\begin{eqnarray}
  {\cal L}_U = \frac{1}{4 g^2} \,F_{\mu\nu}^2 + 
  |D_\mu{\bm\phi}|^2 +  r\, \bar{\bm\phi} \cdot {\bm\phi} +
u \,(\bar{\bm\phi} \cdot {\bm\phi})^2 ,
\label{SAHFT}
\end{eqnarray}
where $F_{\mu\nu}\equiv \partial_\mu A_\nu - \partial_\nu A_\mu$ and
$D_\mu \equiv \partial_\mu + i A_\mu$. Beside the Abelian $U(1)$ gauge
invariance, the UAHFT has a global $SU(N)$ symmetry, ${\bm \phi} \to V
{\bm \phi}$ with $V\in SU(N)$.  The UAHFT is expected to describe the
critical behavior of lattice models with analogous properties, i.e., a
$U(1)$ gauge invariance and an $SU(N)$ global symmetry, but only when
the $U(1)$ gauge field develops critical correlations at the
transition~\cite{BPV-22}. Transitions where the gauge fields play only
the role of hindering the non-gauge-invariant modes from becoming
critical are not described by a gauge field-theory model; they
generically admit a description in terms of a gauge-invariant bilinear
scalar order parameter only~\cite{PW-84,PV-19-AH3d,BPV-19,BPV-22}.

$SU(N)$-symmetric lattice AH models (ULAHM) have been extensively
investigated, using compact and noncompact gauge fields, see, e.g.,
Refs.~\cite{FMS-81,CFIS-05,PV-19-AH3d,BPV-20-hcAH} and
Refs.~\cite{FS-79,DH-81,KK-85,BN-86,BN-86-b,SSSNH-02,SSNHS-03,
  BPV-21-ncAH,BPV-22,BPV-23c}, respectively. In particular, a
noncompact lattice formulation of the three-dimensional (3D) ULAHM is
obtained by considering $N$-component unit-length complex vectors
${\bm z}_{\bm x}$ (satisfying $\bar{\bm z}_{\bm x} \cdot {\bm z}_{\bm
  x} =1$) defined on the sites of a cubic lattice, noncompact gauge
variables $A_{{\bm x},\mu}\in {\mathbb R}$ ($\mu=1,2,3$) defined on
the lattice links, and the nearest-neighbor Hamiltonian
\begin{eqnarray}
H_U = \frac{\kappa}{2} \sum_{{\bm x},\mu>\nu} F_{{\bm x},\mu\nu}^2 -
2NJ \sum_{{\bm x},\mu} {\rm Re}\,( \lambda_{{\bm x},\mu} \bar{\bm
  z}_{\bm x} \cdot {\bm z}_{{\bm x}+\hat\mu}),\quad
\label{SLAH}
\end{eqnarray}
where $\hat{\mu}$ are the lattice unit vectors, and
\begin{eqnarray}
F_{{\bm
    x},\mu\nu}\equiv\Delta_{\mu} A_{{\bm x},\nu} - \Delta_{\nu}
  A_{{\bm x},\mu}, \quad
\lambda_{{\bm x},\mu} \equiv e^{iA_{{\bm x},\mu}},
\label{fdef}
\end{eqnarray}
where $\Delta_\mu A_{{\bm x},\nu} = A_{{\bm x}+\hat{\mu},\nu}- A_{{\bm
    x},\nu}$.  The ULAHM has a local $U(1)$ gauge invariance, ${\bm
  z}_{\bm x} \to e^{i\Lambda_{\bm x}} {\bm z}_{\bm x}$ and $A_{{\bm
    x},\mu} \to A_{{\bm x},\mu} + \Lambda_{\bm x}-\Lambda_{{\bm
    x}+\hat{\mu}}$ with $\Lambda_{\bm x}\in \mathbb{R}$, and a global
$SU(N)$ symmetry ${\bm z}_{\bm x} \to V{\bm z}_{\bm x}$ with $V\in
SU(N)$.

In this work we consider extensions of the $SU(N)$-symmetric AH
  field theories, such as those already introduced in
  Ref.~\cite{MU-90}, obtained by adding scalar gauge-invariant
  self-interactions that are not invariant under $SU(N)$
  transformations, but only under a smaller symmetry group.  We
require, however, the global symmetry group to be such that
$\bar{\bm\phi}\cdot {\bm \phi}$ is the only quadratic combination of
the scalar fields that is invariant under the global and local
symmetry transformations. This requirement guarantees that the
effective field theory describes a standard critical behavior
\cite{ZJ-book,PV-02}, which can be observed by tuning a single
Hamiltonian parameter.  Indeed, in the presence of two or more
quadratic terms that are invariant under the global symmetry group,
the effective field theory describes a multicritical
behavior~\cite{LF-72,FN-74,NKF-74,PV-02,CPV-03}.  In the context of
scalar theories without gauge fields this issue has been extensively
studied, considering Landau-Ginzburg-Wilson (LGW) $\Phi^4$ with
complex symmetry-breaking
patterns~\cite{Aharony-76,ZJ-book,PV-02,v-07}.

To be concrete, in most of the paper we will consider the simplest
breaking of the $SU(N)$ global symmetry, adding a term
$|{\bm\phi}\cdot {\bm\phi}|^2$ to the Lagrangian (\ref{SAHFT}).  We
consider therefore the $SO(N)$-symmetric AH field theory (OAHFT) with
Lagrangian
\begin{eqnarray}
  &&{\cal L}_O = \frac{1}{4 g^2} \sum_{\mu\nu} F_{\mu\nu}^2 + 
   \sum_{\mu} |D_\mu{\bm\phi}|^2 +
  r\, \bar{\bm\phi} \cdot {\bm\phi} + V_O({\bm \phi}),\;\;\quad
  \label{OAHFT}\\
  && V_O ({\bm \phi})=  u \,(\bar{\bm\phi} \cdot
  {\bm\phi})^2 + v \,|{\bm\phi}\cdot {\bm\phi}|^2,
\label{ONpotential}
\end{eqnarray}
with $u\ge 0$ and $u+v\ge 0$, to guarantee the stability of the
potential.  The added term breaks the global $SU(N)$ symmetry, making
the theory invariant only under $SO(N)$ transformations. However, the
symmetry group is still large enough to guarantee that $\bar{\bm\phi}
\cdot {\bm\phi}$ is the only quadratic invariant allowed by the global
and local symmetries.  For $N=1$ the two terms in $V_O$ are
equivalent, thus one recovers the standard one-component AH field
theory.

An analogous extension can be considered for the ULAHM (\ref{SLAH}),
considering the 3D $SO(N)$-symmetric lattice AH model (OLAHM) defined by
the partition function
\begin{eqnarray}
Z_O  &=& \sum_{\{{\bm z},{\bm A}\}} e^{-H_O({\bm z},{\bm A})},\label{OLAHZ}\\
H_O &=& H_U + v \,\sum_{\bm x}|{\bm z}_{\bm
  x}\cdot {\bm z}_{\bm x}|^2.
\label{OLAH}
\end{eqnarray}
One can easily check that the OAHFT (\ref{OAHFT}) corresponds to the
formal continuum limit of the OLAHM with $\kappa=g^{-2}$, after
relaxing the unit-length constraint for the scalar field.

We show that some qualitative features of the phase diagram are the
same in the ULAHM and OLAHM: in both models three different phases
occur, that differ in the properties of the gauge correlations, in the
confinement or deconfinement of the charged excitations, and in the
behavior under the global symmetry transformations. However, the
ordered phases, in particular the Higgs phase, and the nature of the
transition lines crucially depend on the global symmetry-breaking
pattern that is determined by the specific form of the scalar
self-interaction potential.

  We remark that the study of the phase diagrams and critical
  behaviors of extensions of the AH gauge theories, allowing for more
  general scalar potentials and different global symmetry groups, may
  lead to a more thorough understanding of the possible critical
  behaviors that can be observed in the presence of an emergent
  Abelian gauge symmetry.

The paper is organized as follows. In Sec.~\ref{oahftsec} we study the
possible symmetry-breaking patterns and investigate the RG flow of the
OAHFT (\ref{OAHFT}) close to four dimensions.  In Sec.~\ref{olahsec}
we discuss the phase diagram and the nature of the phase transitions
of the OLAHM (\ref{OLAH}).  Finally, in Sec.~\ref{conclu} we draw our
conclusions, and briefly discuss further extensions of the scalar
potential. 

\section{$SO(N)$-symmetric Abelian Higgs field theories}
\label{oahftsec}

\subsection{Global symmetry breaking patterns and order parameters}
\label{sec.2A}

As in the ULAHM, also in $SO(N)$ invariant models we expect
transitions characterized by the breaking of the global symmetry. It
is therefore important to determine the possible symmetry-breaking
patterns.  This analysis can be performed in the mean-field
approximation, since space fluctuations are only relevant along the
transition lines. To characterize the different phases within the
mean-field framework, we need to determine the minima of the effective
Hamiltonian
\begin{equation}
   H_{\rm mf} = r \bar{\bm \phi}\cdot {\bm \phi} + V_O({\bm \phi}).
\label{HMF}
\end{equation}
The analysis is straightforward. For $r> 0$, the minimum corresponds
to ${\bm \phi} = 0$, so that, for $r > 0$, the system is in the
disordered phase in which the symmetry is unbroken. For $r < 0$, we
find two nontrivial sets of minima.
For $-u < v < 0$, the minimum of $H_{\rm mf}$ is obtained for
\begin{equation}
  \bm\phi = e^{i\alpha} \bm{s}, 
\label{minimum1}
\end{equation}
where $\bm{s}$ is a real $N$-component vector, and $\alpha$ an
arbitrary phase.  For $v > 0$, the fields corresponding to the minimum
configurations can be parametrized as
\begin{equation}
  \bm{\phi} = {1\over \sqrt{2}}\left( \bm{s}_1 + i \bm{s}_2\right), \qquad
  \bm{s}_1 \cdot \bm{s}_2 = 0,
\label{minimum2}
\end{equation}
where $\bm{s}_1$ and $\bm{s}_2$ are orthogonal real vectors satisfying
$|\bm{s}_1| = |\bm{s}_2|$, so that
$\bar{\bm\phi}\cdot{\bm\phi} = s^2 \equiv s_1^2 =s_2^2$.

Using these results we can determine the global symmetry of the broken
phases. For $v < 0$, the broken phase is invariant under $O(N-1)$
transformations, while for $v > 0$ the residual symmetry is
$SO(2)\otimes O(N-2)$.  Note that the $SO(2)$ subgroup for $v>0$
corresponds to the transformations that rotate the two vectors ${\bm
  s}_1$ and ${\bm s}_2$ in the plane in which they lie together with a
change of phase.  To write these transformations explicitly, let us
note that, by an appropriate $SO(N)$ rotation, we can always take
${\bm \phi} = (A,\pm iA,0,\ldots,0)$. It is then immediate to verify
that the vector ${\bm \phi}$ is left invariant by the transformation
\begin{equation}
\phi^a \to e^{\mp i\alpha} \sum_{ab} V^{ab} \phi^b, 
\label{SO2symmetry}
\end{equation}
where 
$V = V_2 \oplus I_{N-2}$, $I_{N-2}$ is the $(N-2)$-dimensional 
identity matrix and 
\begin{equation}
V_2 = \begin{pmatrix} 
       \cos\alpha & \sin\alpha \\
       -\sin\alpha & \cos\alpha 
    \end{pmatrix}   .
\end{equation}
We should also consider discrete transformations, as they can also
play a role at the transitions. The Lagrangian (\ref{OAHFT})
and the mean-field Hamiltonian (\ref{HMF}) are invariant under the
${\mathbb Z}_2$ transformation ${\bm \phi} \to \bar{\bm \phi}$. 

Because of gauge invariance, minimum configurations that differ by
gauge transformations are equivalent. For negative $v$, we can
therefore set $\alpha = 0$ in Eq.~(\ref{minimum1}) and consider only
real vectors.  The invariance group of the broken phase is ${\mathbb
  Z}_2\otimes O(N-1)$, the ${\mathbb Z}_2$ transformations
corresponding to ${\bm \phi} \to \bar{\bm \phi}$.  For positive values
of $v$, if we apply a gauge transformation to the minimum
configurations, we obtain ${\phi'}^a = {s'}_1^a + i {s'}_2^a$ with
\begin{eqnarray}
 {s'}_1^a &=& \cos \alpha\ s_1^a + \sin \alpha\ s_2^a ,\nonumber \\
 {s'}_2^a &=& -\sin \alpha\ s_1^a + \cos \alpha\ s_2^a .
\label{rotations-s1s2}
\end{eqnarray}
We can thus freely rotate the two vectors in the plane in which they
lie.  This implies that, modulo gauge transformations, minimum
configurations are classified by the relative orientation of the two
vectors (chirality) and by the plane in which ${\bm s}_1$ and ${\bm
  s}_2$ lie. The $SO(2)$ symmetry (\ref{SO2symmetry}) is irrelevant as
it also involves a gauge transformation, so that the invariance group
of the ordered phase is $O(N-2)$.

We wish now to identify appropriate order parameters for the two
different symmetry-breaking patterns. In the UAHFT an order-parameter
field is provided by the complex gauge-invariant bilinear operator
\begin{equation}
  Q^{ab}({\bm x}) = \bar{\phi}^a({\bm x}) \phi^b({\bm x}) - {1 \over
      N} \bar {\bm \phi}({\bm x})\cdot {\bm \phi}({\bm x})\, \delta^{ab}.
  \label{qoperft}
\end{equation}
We now show that 
\begin{eqnarray}
  R^{ab}({\bm x}) = {\rm Re}\,Q^{ab}({\bm x}),\qquad
  T^{ab}({\bm x}) = {\rm Im}\,Q^{ab}({\bm x}),
  \label{wqoperft}
\end{eqnarray}
which transform under different representations of the $SO(N)$ group,
provide the order-parameter fields for the $SO(N)$ theory. Indeed, if
the minimum configurations are given by Eq.~(\ref{minimum1}), thus for
$v<0$, we have $T^{ab} = 0$ and
\begin{equation}
   R^{ab} = s^a s^b - {s^2\over N} \delta^{ab},
\end{equation}
where $s=|{\bm s}|$.  Instead, if the minimum configurations are given
by Eq.~(\ref{minimum2}), thus for $v>0$, we have
\begin{eqnarray}
  R^{ab} &=& {1\over 2}\left(
  s_1^a s_1^b + s_2^a s_2^b\right) - { s^2 \over N} \delta^{ab},
\label{rtmin}\\
T^{ab} &=& {1\over 2}\left( s_1^a s_2^b - s_1^b  s_2^a\right),
\nonumber
\end{eqnarray}
where $s^2 = |\bm{s}_1|^2 = |\bm{s}_2|^2$.  Note that only the order
parameter $T^{ab}$ is sensitive to the breaking of the ${\mathbb Z}_2$
symmetry ${\bm \phi}\to \bar{\bm \phi}$. Moreover, Eq.~(\ref{rtmin})
implies the relation
\begin{equation}
(T^2)^{ab} = -
  {s^2\over 2} R^{ab} - {s^4\over 2N} \delta^{ab},\qquad
  {\rm Tr}\,T^2 = - {s^4\over 2} .
  \label{trrel}
  \end{equation}
In particular, for $N = 2$ we have that $R^{ab} = 0$ (keeping into
accont that $\bm{s}_1\cdot\bm{s}_2=0$), and we can write $T^{ab} =
\frac{1}{2} \sigma s^2 \epsilon^{ab}$, where $\epsilon^{ab}$ is the
two-index antisymmetric tensor ($\epsilon^{12}=-\epsilon^{21}=1$ and
$\epsilon^{11}=\epsilon^{22}=0$), and $\sigma$ is a variable that
takes only the values $\pm 1$, which is related to the relative
orientation (chirality) of the orthogonal pair $({\bm s}_1, {\bm
  s}_2)$.

The analysis of the possible phases could have also been performed in
the real formulation of the model. Indeed, the Lagrangian
(\ref{OAHFT}) can also written in terms of a real matrix field
$\varphi_{ai}$ ($a=1,...,N$ and $i=1,2$) defined by $\phi_a =
\varphi_{a1} + i \varphi_{a2}$. We obtain the equivalent Lagrangian
\begin{eqnarray}
&&{\cal L} = {1 \over 4 g^2} \sum_{\mu\nu} F^2_{\mu\nu} + 
\sum_{\mu,ai} (D_\mu \varphi)_{ai}^2 + 
  r\sum_{ai}  \varphi_{ai}^2  
  \label{LGWvarphi}\\
&&\;\;+ \tilde{u} \,\Bigl( \sum_{ai} \varphi_{ai}^2\Bigr)^2  +
\tilde{v} \,\Bigl[ \sum_{ij} \Bigl(\sum_a \varphi_{ai} \varphi_{aj}\Bigr)^2 -
  \Bigl(\sum_{ai} \varphi_{ai}^2\Bigr)^2\Bigr],
  \nonumber
\end{eqnarray}
where $\tilde{v}=2v$ and $\tilde{u} = u + v$, $(D_\mu \varphi)_{ai} =
\partial_\mu \varphi_{ai} - i A_\mu \epsilon_{ab} \varphi_{bi}$.  Of
course, the results for the symmetry-breaking patterns reported in
Ref.~\cite{CPPV-04} are equivalent to those reported above.  The
real-field formulation with Lagrangian (\ref{LGWvarphi}) shows that
the OAHFT can be interpreted as the Abelian gauge theory obtained by
gauging the $SO(2)$ group of the global symmetry group $O(2)\otimes
O(N)$.

\subsection{RG flow and fixed points}
\label{fpsec}

Let us now discuss the RG flow of the OAHFT.  Its main features
  can be already inferred from earlier field-theoretical perturbative
  analyses~\cite{Hikami-80,MU-90}, in particular, from the one-loop
  perturbative computations within the minimal subtraction
  renormalization scheme of the dimensional regularizaton, reported in
  Ref.~\cite{Hikami-80}. As we shall see, they indicate that, as it
  occurs for the UAHFT, the RG flow of the OAHFT has a stable fixed
  point for a sufficiently large number of scalar components,
  $N>N_O^*(d)$, where $N_O^*(d)$ depends on the dimension $d$ of the
  system.

Close to four dimensions the RG flow can be studied by using the
$\beta$ functions computed in dimensional regularization with the
minimal subtraction renormalization scheme.  Setting $\alpha\equiv
g^2$ and $\varepsilon\equiv 4-d$, the $\beta$ functions associated
with the couplings $u$, $v$ and $\alpha$ are given
by~\cite{Hikami-80}~\footnote{The one-loop series reported in
Eq.~(\ref{betafunc}) can be obtained by some straightforward
manipulations of the one-loop series reported in Ref.~\cite{Hikami-80}
for scalar $O(N)$ gauge theories. Some checks are obtained by noting
that: (i) for $v=0$ one should reobtain the UAHFT
series~\cite{HLM-74}; (ii) for $\alpha = 0$, one should reproduce the
series for the purely scalar model reported in
Refs.~\cite{DPV-04,PV-07}, known up to five loops~\cite{DPV-04,PV-07};
(iii) since the addition of the $v$-term changes the global symmetry
of the model, the $v=0$ plane must be a separatrix of the RG flow,
which implies $\beta_v = - \varepsilon v + v f_v(u,v,\alpha)$; (iv)
since for $N=1$ the two quartic terms are equivalent, for $N\to 1$
the relation $\beta_u(u,s-u,\alpha) + \beta_v(u,s-u,\alpha) =
\beta_u(s,0,\alpha)$ holds.}
\begin{eqnarray}
&&\beta_u = - \varepsilon u + 
  (N+4) u^2 + 4 u v + 4 v^2
  - 18 u \alpha + 54 \alpha^2,
\nonumber \\
&&\beta_v = - \varepsilon v + 
N v^2 + 6 v u -18 v \alpha,\nonumber \\
&&\beta_\alpha = - \varepsilon \alpha + N \alpha^2.  
\label{betafunc}
\end{eqnarray}
The normalizations of the renormalized couplings $u,v,\alpha$ have
been chosen to simplify the formulas (they can be easily inferred from
the above expressions).

Stable fixed points of the RG flow correspond to zeroes of the $\beta$
functions (\ref{betafunc}), such that the eigenvalues of the stability
matrix $\Omega_{ij} = \partial \beta_i/\partial g_j$, computed at the
zero, are all positive. In the large-$N$ limit a stable fixed point
occurs for
\begin{equation}
  \alpha^* = u^* = v^* = {\varepsilon\over N} + O(\varepsilon^2,N^{-2}). 
  \label{lnfixpo}
\end{equation}
The analysis of the $\beta$ functions shows that that this stable
fixed exists for $N>N_O^*(d)$, where $N_O^*(d) = N_{O,4}^* +
O(\varepsilon)$ with $N_{O,4}^* \approx 210$, which is the only
positive solution of the fourth-order equation~\cite{MU-90}
\begin{equation}
n^4 - 204 n^3 - 1356
n^2 - 864 n - 15552 = 0.
\label{no4star}
\end{equation}
For comparison we mention that the UAHFT
has a stable fixed point for $N>N^*(d)$ with $N^*(d) = N_{4}^* +
O(\varepsilon)$ and $N_4^*=94+24 \sqrt{15}\approx 183$.  Therefore,
for $\varepsilon \ll 1$, OAHFT and UAHFT have comparable boundary
values for the existence of a stable fixed point.  In the case of the
UAHFT, $N^*(d)$ significantly decreases approaching $d=3$: one finds
$N^*(d=3)\approx 7$~\cite{BPV-21-ncAH,BPV-22,IZMHS-19,SZJSM-23}.  We
also mention that $N^*(d)$ is expected to converge to a small value
for $d\to 2$~\cite{IZMHS-19}, and that an analogous argument indicates
that $|N_O^*(d\to 2)|=O(1)$ as well.  This suggests that $N_O^*(d)$
has a $d$-dependence similar to $N^*(d)$ for the UAHFT. Therefore, we
may guess that $N_{O}^*(d=3)$ is of order ten, as in the $SU(N)$ case.

It is important to note that the stable fixed point for the OAHFT lies
in the region $v > 0$. Therefore, this fixed point is only relevant
for transitions between a disordered phase and a Higgs phase with a
global $O(N-2)$ symmetry.  On the other hand, on the basis of this RG
analysis, no charged transition is expected if the Higgs phase is
invariant under ${\mathbb Z}_2\otimes O(N-1)$.

The analysis of the RG flow of the OAHFT provides information on the
behavior of the transitions in 3D lattice AH systems, where gauge
fields become critical. If the symmetry of the broken phase is the one
observed in the OAHFT model for $v < 0$, only first-order transitions
are possible. If, instead, the symmetry breaking pattern is the one
observed for $v > 0$, the behavior depends on the number of components
of the scalar field.  For $N>N_O^*(d=3)$, continuous charged
transitions are possible, provided the lattice system is effectively
inside the attraction domain of the stable fixed point. Instead, for
$N< N_O^*(d=3)$, only first-order transitions are possible (unless
there are additional three-dimensional universality classes that are
not analytically related with the four-dimensional RG fixed points, as
it occurs for the one-component AH
models~\cite{BFLLW-96,NRR-03,BPV-23c}).

Beside the stable fixed point, the $\beta$ functions admit other
zeroes with unstable directions. The fixed point with $v^*=0$ that is
stable in the $SU(N)$-symmetric field theory (in the large-$N$ limit
it correspods to $ \alpha^* = u^* \approx {\varepsilon/N}$) is
unstable with respect to the perturbation proportional to $v$, with a
negative eigenvalue $\lambda_v=-\varepsilon +
O(\varepsilon^2,N^{-1})$.  Therefore, the parameter $v$ is a relevant
perturbation of the $SU(N)$-symmetric fixed point, with positive RG
dimension $y_v=-\lambda_v$. The addition of the term proportional to
$v$ drives the flow away from the stable $SU(N)$ fixed point, either
towards the O($N$) stable fixed point (this is only possible for $v >
0$) or towards infinity (in this case first-order transitions are
observed).

We also mention that the RG flow of the scalar model without gauge
fields with Lagrangian (\ref{LGWvarphi}) has been extensively studied,
because it is relevant for the normal-to-planar superfluid transition
in $^3$He~\cite{JLM-76,DPV-04}, and for transitions in some frustrated
magnetic systems with noncollinear
order~\cite{Kawamura-86,Kawamura-88,CPPV-04}.  The RG flow of the
scalar theory close to four dimensions can be inferred from the
analysis of the zeroes of the $\beta$ functions $\beta_u$ and
$\beta_v$, cf. Eq.~(\ref{betafunc}), setting $\alpha=0$.  Close to
four dimensions there is a stable fixed point for $N \gtrsim
22$~\cite{Kawamura-88,PRV-01}. However, the analyses of high-order 3D
perturbative expansions~\cite{DPV-04,CPPV-04} show that, in 3D there
are stable RG fixed points for $N=2$ and $N=3$, that are not connected
with those existing close to four dimensions.

The stable fixed point of the scalar theory without gauge fields is
unstable with respect to the gauge parameter $\alpha\sim \kappa^{-1}$.
A simple analysis of the $\beta$ functions (\ref{betafunc}) shows that
the RG dimension of the gauge perturbation is positive, i.e.,
\begin{equation}
  y_\alpha = - \lambda_\alpha = - \left. {\partial \beta_\alpha \over
    \partial \alpha} \right|_{\alpha=0,u=u^*,v=v^*} = \varepsilon=4-d,
  \label{yalpha}
\end{equation}
where $\lambda_\alpha$ is one of the eigenvalues of the stability
matrix $\Omega_{ij} = \partial \beta_i/\partial g_j$ computed at the
scalar fixed point with $\alpha=0$.  Note that $y_\alpha=4-d$
corresponds to the dimension of the gauge coupling
$\alpha\sim\kappa^{-1}$ in $d$ dimensions.  This result holds to all
order of the $\varepsilon$ expansion, due to the fact that
$\beta_\alpha$ has the general form $\beta_\alpha = - \varepsilon
\alpha + \alpha^2 F(\alpha,u,v)$, where $F(\alpha,u,v)$ has a regular
perturbative expansion~\cite{PV-19-AH3d}.  Therefore, we find
$y_\alpha = 1$ in three dimensions.  Note that the relevance of the
gauge fluctuations at the fixed points of the purely scalar theory,
and therefore the crossover towards a different asymptotic behavior,
is independent of the existence of the stable fixed point of the full
theory, which is only relevant to predict the eventual asymptotic
behavior.

The RG analysis reported above shows that $SO(N)$-symmetric Abelian
gauge systems for large values of $N$ may undergo continuous phase
transitions.  It is interesting to compute the corresponding critical
exponents.  The correlation-length exponent $\nu$ in the large-$N$
limit can be computed by using the results of Ref.~\cite{Hikami-80},
which considered scalar-gauge theories obtained by gauging the $O(M)$
subgroup of the global symmetry $O(M)\otimes O(N)$. Assuming the
existence of a critical transition, the critical exponents for fixed
$M$ and $d$ were computed in the large-$N$ limit.  For $M=2$ and $d =
3$ the correlation-length exponent $\nu$ is given by
\begin{equation}
\nu = 1 - {176 \over 3 \pi^2 N} + O(N^{-2}), 
  \label{nuln}
\end{equation}
which is numerically close to the  large-$N$ result for the
UAHFT~\cite{HLM-74}
\begin{equation}
  \nu = 1 - \frac{48}{\pi^2 N} + O(N^{-2}).
\label{nulnsun}
\end{equation}

\section{Noncompact $SO(N)$-symmetric lattice Abelian Higgs models}
\label{olahsec}

We now discuss the phase diagram and the nature of the phase
transitions in the OLAHM, whose Hamiltonian is given in
Eq.~(\ref{OLAH}).  As we shall see, some qualitative features of the
OLAHM phase diagram are analogous to those of the ULAHM.  However,
substantial changes are expected in the nature of the ordered
low-temperature phases, in particular of the Higgs phase, and of the
transition lines, which crucially depend on the global
symmetry-breaking pattern that, in turn, depends on the specific form
of the scalar self-interaction potential.

\subsection{$SU(N)$-symmetric lattice Abelian-Higgs models}
\label{sunsymmnonc}

\begin{figure}[tbp]
\includegraphics*[width=0.95\columnwidth]{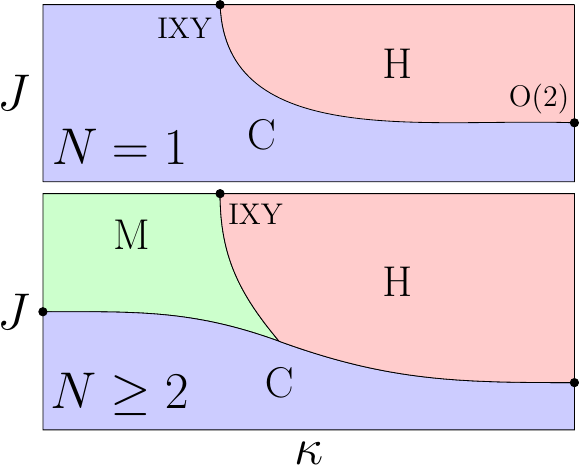}
  \caption{The $\kappa$-$J$ phase diagram of lattice AH models, for
    $N=1$ (top) and generic $N\ge 2$ (bottom).  For $N\ge 2$
    three phases are present: the small-$J$ Coulomb
    (C) phase, in which the scalar field is disordered and gauge
    correlations are long ranged; the large-$J$ molecular (M) and
    Higgs (H) ordered phases, in which the global symmetry is
    spontaneously broken.  For $N=1$ there are only two phases: the
    Coulomb and the Higgs phase. 
  }
\label{phadiaLAH}
\end{figure}

We begin by reviewing what is known for the $SU(N)$-symmetric ULAHM.  A
sketch of the $\kappa$-$J$ phase diagram of the ULAHM is shown in
Fig.~\ref{phadiaLAH}. For $N\ge 2$ three phases occur. In the
small-$J$ Coulomb (C) phase, the scalar field is disordered and gauge
correlations are long ranged.  For large $J$ two phases occur, the
molecular (M) and the Higgs (H) ordered phase, in which the global
symmetry is spontaneously broken from $SU(N)$ to $U(N-1)$, with the
emergence of $2N-2$ long ranged Goldstone modes~\cite{MZ-03}.  An
appropriate order parameter is the gauge-invariant bilinear operator
\begin{equation}
  Q_{L,\bm x}^{ab} = \bar{z}_{\bm x}^a z_{\bm x}^b - {1\over N}\delta^{ab},
  \label{qoper}
\end{equation}
which is the lattice analogue of the field-theory bilinear operator
(\ref{qoperft}). For $N=1$ there is no global symmetry, therefore
there are only two phases, see Fig.~\ref{phadiaLAH}, the Coulomb phase 
and the Higgs phase, that are characterized by the behavior of
nonlocal gauge-invariant charged operators~\cite{BPV-23b,KK-85,KK-86,BN-87}, 
which are confined in the former and deconfined in the latter.

The two ordered phases of multicomponent systems are distinguished by
the behavior of the gauge modes: the gauge field is long ranged in the
M phase (small $\kappa$), while it is gapped in the H phase (large
$\kappa$).  Moreover, while the C and M phases are confined phases, in
the H phase charged excitations, represented by gauge-invariant
nonlocal dressed scalar
operators~\cite{Dirac:1955uv,KK-85,KK-86,BPV-23b}, are
deconfined~\cite{BPV-23b,KK-85,KK-86,BN-87}.  The transition lines may
be of first order or continuous and, in the latter case, belong to
universality classes that may depend on the number $N$ of scalar
components.  The continuous transitions are related with the stable
(charged or uncharged) fixed points of the RG flow, each one with its
own attraction domain in the model parameter space.

The CH and CM transitions are both characterized by the spontaneous
breaking of the global $SU(N)$ symmetry, but differ in the role of the
gauge fields.  The continuous transitions along the CH line are
charged transitions, where gauge fields become critical, and are
associated with the stable fixed point of the RG flow of the UAHFT
(\ref{SAHFT}). As already mentioned in Sec.~\ref{fpsec}, continuous
transitions can only be observed for $N>N^\star$ with $N^\star=
7(2)$~\cite{BPV-21-ncAH}.  At the CM transitions gauge fields play no
role (gauge correlations are long ranged on both sides of the
transition) and thus an effective description can be obtained by
considering an $SU(N)$ symmetric LGW theory defined in terms of the
complex hermitian field $\Psi^{ab}$, that corresponds to
the bilinear operator $Q^{ab}$, without gauge fields
\cite{PV-19-CP,BPV-21-ncAH}. This predicts that continuous transitions
occur only in two-component systems, i.e., for $N=2$. Their critical
behavior belongs to the O(3) vector (Heisenberg) universality
class~\cite{ZJ-book,PV-02}.

While transitions along the CM and CH lines are related with the
spontaneous breaking of the global symmetry, the MH line separates two
ordered phases that differ only in the behavior of the gauge
correlations, without a local gauge-invariant order parameter.  The
continuous transitions along the MH line belong to the same
universality class as those in the inverted $XY$ ($IXY$)
model~\cite{BPV-23c}, which is related with the standard $XY$ model with
Villain action by a duality transformation~\cite{NRR-03}. This is the
same universality class controlling the topological critical behavior
of the CH transitions in the one-component lattice AH
model~\cite{BPV-23c}. Note that this apparently simple behavior along
the MH transition line is not obvious, because of the simultaneous
presence of the massless gauge modes that drive the $IXY$ transitions
and of the long ranged (zero-mass) Goldstone bosons, related with the
breaking of the global $SU(N)$ symmetry.  The numerical analyses
reported in Ref.~\cite{BPV-23c} show that, along the MH line, the
massless Goldstone modes effectively decouple from the massless gauge
modes that drive the critical behavior, so that the finite-$J$
transitions belong to the $IXY$ universality class.

\subsection{$SO(N)$-symmetric lattice Abelian Higgs models}
\label{sonsymmnonc}

We wish now to discuss the general features of the phase diagram the
OLAHM. We will argue that the phase diagram is similar to that of the
ULAHM, shown in Fig.~\ref{phadiaLAH}.  Indeed OLAHM presents three
phases as well: one small-$J$ disordered phase and two large-$J$
phases in which the global $SO(N)$ symmetry is spontaneously broken.
As in the ULAHM, the ordered phases differ in the confinement
properties of the nonlocal charged
excitations~\cite{KK-86,BN-87,BPV-23b,BPV-23c}, and in the nature of
the gauge modes.

To characterize the spontaneous breaking of the $SO(N)$ symmetry, we
consider the lattice analogue of the field-theory operators defined in
Eq.~(\ref{wqoperft}),
\begin{eqnarray}
 R_{L,\bm x}^{ab} &=& {\rm Re}\,Q_{L,\bm x}^{ab} = {1\over 2}(\bar{z}_{\bm
    x}^a z_{\bm x}^b + \bar{z}_{\bm x}^b z_{\bm x}^a) - {1\over
    N}\delta^{ab},
  \label{wqoper} \\
  T_{L,\bm x}^{ab} &=& 
   {\rm Im}\,Q_{L,\bm x}^{ab} = {1\over 2i} (\bar{z}_{\bm x}^a
  z_{\bm x}^b - \bar{z}_{\bm x}^b z_{\bm x}^a) ,
  \label{wqopera}
\end{eqnarray}
which transform under two different irreducible representations of the
$SO(N)$ group.  As already discussed, in the disordered phase both
order parameters vanish.  Since the symmetry-breaking pattern depends
on the sign of $v$, the behavior of $R_{L,\bm x}^{ab}$ and $T_{L,\bm
  x}^{ab}$ depends on the sign of $v$.  For $v < 0$, $R_{L,\bm
  x}^{ab}$ condenses in the ordered phase, while $T_{L,\bm x}^{ab}$
still vanishes.  For $v > 0$ and $N =2$, $T_{L,\bm x}^{ab}$ condenses,
while $R_{L,\bm x}^{ab}$ is constant. Finally, for $v > 0$ and $N \ge
3$, both order parameters condense in the ordered phase.

\subsubsection{The MH transition line at low-temperature}

The existence of two different large-$J$ ordered phases which differ
for the topological properties of the gauge field---charged
excitations are confined/deconfined in the M and H phase,
respectively---is suggested by the existence of a transition point for
$J=\infty$. The argument is the same that holds for the ULAHM.  For
$J\to\infty$, the relevant configurations are those that maximize
$\sum_{{\bm x},\mu} {\rm Re}\,( \bar{\bm z}_{\bm x} \cdot
\lambda_{{\bm x},\mu}\,{\bm z}_{{\bm x}+\hat\mu})$, independently of
the scalar potential.  This implies ${\bm{z}}_{\bm x} = \lambda_{{\bm
    x},\mu}\, {\bm z}_{{\bm x}+\hat\mu}$, and therefore $\lambda_{{\bm
    x},{\mu}} \,\lambda_{{\bm x}+\hat{\mu},{\nu}}
\,\bar{\lambda}_{{\bm x}+\hat{\nu},{\mu}} \,\bar{\lambda}_{{\bm
    x},{\nu}} = 1$ for each lattice plaquette. Then, by an appropriate
gauge transformation, we can set $A_{{\bm x}, \mu} = 2 \pi n_{{\bm x},
  \mu}$, where $n_{{\bm x}, \mu} \in {\mathbb Z}$, obtaining the $IXY$
model, which has a transition in the $XY$ universality class, at
$\kappa_c = 0.076051(2)$~\cite{NRR-03,BPV-21-ncAH} (estimates of 
critical exponents can be found in 
Refs.~\cite{CHPV-06,Hasenbusch-19,CLLPSSV-20}). 
Therefore, for
$J\to \infty$ the OLAHM has a transition for any value of $v$.

A natural hypothesis is that the $J\to\infty$ $IXY$ transition point
is the starting point of a finite-$J$ line (MH line) of transitions,
whose nature, as in the ULAHM~\cite{BPV-23c}, is independent of $J$,
at least for sufficiently large finite $J$ (we cannot exclude that the
transitions turn into first-order ones before the multicritical point, 
where the three
transition lines meet). The transitions should belong to the $IXY$
universality class, as the MH transitions in the ULAHM and the CH
transitions in the one-component lattice AH model, see
Fig.~\ref{phadiaLAH}.  This universal behavior of the MH transitions
is possible if, as observed in the ULAHM, the gauge critical modes
driving the $IXY$ transitions decouple from the long-range Goldstone
modes present in the ordered phases.  If this occurs, the scalar
degrees of freedom are irrelevant and so are the global symmetry and
the symmetry-breaking pattern.

\subsubsection{The CM transition line}

As in the ULAHM, we expect transitions along the CM line, i.e., for
small values of $\kappa$, to have the same universal features as those
occurring for $\kappa=0$.  For $\kappa=0$ the gauge variables can be
integrated out in the partition function (\ref{OLAHZ}), obtaining the
model with Hamiltonian
\begin{eqnarray}
  H_{O,\kappa=0} &=& - \sum_{{\bm x},\mu} \ln\,I_0(2NJ
  |\bar{\bm
    z}_{\bm x} \cdot {\bm z}_{{\bm x}+\hat\mu}|)\nonumber\\
  &+& v \sum_{\bm x} \,|{\bm z}_{\bm x}\cdot {\bm z}_{\bm x}|^2,
    \label{LAHHk0}
\end{eqnarray}
where $I_0(x)$ is the modified Bessel function. We recall that
$I_0(x)=I_0(-x)$, $I_0(x) = 1 + x^2/4 + O(x^4)$, and $I_0(x)\approx
e^x/\sqrt{2\pi x}$ for large $x$.  In the absence of the $v$-term,
this Hamiltonian provides a lattice fomulation of the $CP^{N-1}$
model~\cite{DHMNP-81}, which is equivalent, as far as the critical behavior 
is concerned,  to the standard one with Hamiltonian
\begin{equation}
H_{CP} = - J N \sum_{{\bm x},\mu} |\bar{\bm z}_{\bm x} \cdot
{\bm z}_{{\bm x}+\hat\mu}|^2. 
\label{standardCPN}
\end{equation}
For $\kappa=0$, and for sufficiently small values of $\kappa$ along
the CM transition line, gauge fluctuations are not expected to play an
active role at the transition. Indeed, the gauge properties of the C
and M phases are the same: gauge modes are long ranged and charged
excitations are confined in both of them.  Therefore, the transition
should be uniquely driven by the breaking of the global
symmetry. Therefore, we expect that an effective description of the
critical universal behavior can be obtained by considering a LGW
theory for the gauge-invariant scalar order parameter that condenses
at the transition, without considering the gauge fields
~\cite{PV-19-CP,PV-19-AH3d,BPV-21-ncAH}.

For $v < 0$ the relevant order parameter is $R_{L,\bm x}^{ab}$, which
is a real symmetric operator. Therefore, we expect the small-$\kappa$
transitions along the CM line to be described by a LGW for a real
symmetric traceless $N\times N$ matrix field $\Phi^{ab}({\bm x})$,
that represents a coarse-grained average of $R_{L,\bm x}^{ab}$ over a
large, but finite, lattice domain.  The corresponding LGW Lagrangian
is obtained by considering all monomials in $\Phi^{ab}({\bm x})$ that
are allowed by the global $SO(N)$ symmetry up to fourth order.  We
obtain
\begin{eqnarray}
  {\cal L} &=& {\rm Tr} (\partial_\mu \Phi)^2 + r \,{\rm Tr} \,\Phi^2
  + s \,{\rm tr} \,\Phi^3
\label{hlg}   \\
&+&   \,u \, ({\rm Tr}
\,\Phi^2)^2 + v\, {\rm Tr}\, \Phi^4.
\nonumber
\end{eqnarray}
For $N=2$, the cubic term vanishes and the two quartic terms are
equivalent.  The resulting LGW theory is equivalent to that of the
$O(2)$-symmetric vector model, thus predicting that continuous
transitions belong to the $XY$ universality class.  On the other hand,
for $N\ge 3$ the cubic $\Phi^3$ term is generally present. This is
usually considered as the indication that phase transitions occurring
in systems sharing the same global properties are of first order, as
one can easily infer using mean-field arguments.  We expect this
behavior to hold fon any $v<0$, up to $v=0$ where the $SU(N)$ symmetry
is restored, and we recover the $CP^{N-1}$ model, whose transition is
continuous for $N=2$, in the $O(3)$ vector universality class, and of
first order for any $N\ge 3$ \cite{PV-19-CP,PV-20-largeNCP}.

We can explicitly verify the above predictions by considering the
limit $v \to -\infty$. In this limit the relevant configurations are 
those that minimize the potential $V_O$, as discussed in 
Sec.~\ref{sec.2A}. Indeed, we should again minimize $H_{\rm MF}$ where 
$r$ now plays the role of the Lagrange multiplier that enforces the 
condition $\bar{\bm z} \cdot {\bm z} = 1$.
For $v < 0$, we have 
\begin{equation}
   {\bm z}_{\bm x} = e^{i\theta_{\bm x}} {\bm s}_{\bm x},
  \label{parametrization-z-vneg}
\end{equation}
where ${\bm s}_{\bm x}$ is a real unit-length $N$-component vector.
This representation is redundant as the pair ${\bm s}_{\bm x}$,
$\theta_{\bm x}$ and the pair ${\bm s}_{\bm x}' = - {\bm s}_{\bm x}$,
$\theta_{\bm x}' = \theta_{\bm x} + \pi$ both correspond to ${\bm
  z}_{\bm x}$.  Thus, this parametrization of the scalar field
introduces an additional kinematical ${\mathbb Z}_2$ gauge
invariance. Using the parametrization (\ref{parametrization-z-vneg}),
the scalar hopping term becomes
\begin{equation}
 \bar{\bm z}_{\bm x} \cdot {\bm z}_{{\bm x} + \hat{\mu}} 
    = e^{i (\theta_{{\bm x} + \hat{\mu}} - \theta_{{\bm x}}) }\,
  {\bm s}_{\bm x} \cdot {\bm s}_{{\bm x} + \hat{\mu} }.
 \label{ZbarZvneg}
\end{equation}
If we choose the unitary gauge, we can set $\theta_{\bm x} = 0$,
obtaining a theory in terms of $A_{{\bm x},\hat \mu}$ and ${\bm
  s}_{\bm x}$, which is invariant under ${\mathbb Z}_2$ gauge
transformations that involve both ${\bm s}_{\bm x}$ and 
$A_{{\bm x},\mu}$.  Substituting Eq.~(\ref{ZbarZvneg}) into the
$\kappa=0$ Hamiltonian (\ref{LAHHk0}), we obtain
\begin{eqnarray}
  H &=& - \sum_{{\bm x},\mu} \ln\,I_0(2NJ
  |{\bm
    s}_{\bm x} \cdot {\bm s}_{{\bm x}+\hat\mu}|).
\end{eqnarray}
This Hamiltonian is invariant under global $O(N)$ and local ${\mathbb
  Z}_2$ transformations and is ferromagnetic (for large $J$ the fields
${\bm s}_{\bm x}$ order). Thus, it represents a variant Hamiltonian of
the so-called $RP^{N-1}$ model, which is characterized by a global
$O(N)$ invariance and a local ${\mathbb Z}_2$ gauge symmetry.
This
result is consistent with the LGW description given above. Indeed, in
the $RP^{N-1}$ models the relevant order parameter is
\begin{equation}
 S_{\bm x}^{ab} = s^a_{\bm x} s^b_{\bm x} - {1\over N} \delta^{ab},
\end{equation}
so that the LGW fundamental field is a real symmetric traceless tensor
$\Phi^{ab}$. The corresponding Lagrangian is given in Eq.~(\ref{hlg}).

The behavior changes when we consider the opposite case $v > 0$. To
understand some general features of the critical behavior, we begin by
studying the lattice model in the limit $v\to \infty$.  The relevant
configurations are those that minimize the potential $V_O$ for $v>0$,
see Sec.~\ref{sec.2A}.  Since $|{\bm z}_{\bm x}| = 1$, they can be
written as
\begin{equation}
  {\bm z}_{\bm x} = {1\over \sqrt{2}} ({\bm s}_{1,{\bm x}} +
  i {\bm s}_{2,{\bm x}}),\qquad {\bm
    s}_{1,{\bm x}}\cdot{\bm s}_{2,{\bm x}}=0,
  \label{vinfconf}
\end{equation}
where ${\bm s}_1$ and ${\bm s}_2$ are two real orthogonal unit-length
vectors (${\bm s}_1^2 = {\bm s}_2^2 =1$ so that $\bar{\bm z}\cdot {\bm
  z}=1$).  One can easily check that the $\kappa=0$ Hamiltonian
(\ref{LAHHk0}) can be written in terms of the antisymmetric tensor
field $T^{ab}_{L,\bm x}$ only. Indeed, since $Q_{L,{\bm x}} =
R_{L,{\bm x}} + i T_{L,{\bm x}}$ and
\begin{eqnarray}
  && |\bar{\bm z}_{\bm x} \cdot {\bm z}_{{\bm x}+\hat\mu}|^2 =
  {\rm Tr} \, Q_{L,\bm x} Q_{{L,\bm
      x}+\hat\mu}  + {1\over N}  \nonumber \\
  && \;\;=
  {\rm Tr} \, R_{L,\bm x} R_{{L,\bm
      x}+\hat\mu} 
   - {\rm Tr} \, T_{L,\bm x} T_{L,{\bm x}+\hat\mu} 
    + {1\over  N} ,\qquad
\label{userel}
\end{eqnarray}
using Eq.~(\ref{trrel}) with $s^2=1$ we obtain
\begin{eqnarray}
  |\bar{\bm z}_{\bm x} \cdot {\bm z}_{{\bm x}+\hat\mu}|^2 = 
    - {\rm Tr} \, T_{L,\bm x} T_{L,{\bm x}+\hat\mu} 
    + 4 {\rm Tr} \, T_{L,\bm x}^2  
     T_{L,{\bm x}+\hat\mu}^2 ,
     \label{HTT}
\end{eqnarray}
This expression drastically
simplifies for $N=2$ and $N=3$. For $N=2$, the only relevant degree of
freedom is the chirality of the configuration, which can be expressed
in terms of the variable $\sigma_{\bm x} = \sum_{ab} \epsilon^{ab}
s^a_{1,\bm x} s^b_{2,\bm x}$, that can only take the values $\pm 1$.
As expected, the gauge-invariant tensor $T^{ab}_{L,\bm x}$ depends
only $\sigma_{\bm x}$: $T^{ab}_{L,\bm x} = \frac{1}{2}\epsilon^{ab}
\sigma_{\bm x}$. Substituting in Eq.~(\ref{HTT}) we obtain
\begin{equation}
|\bar{\bm z}_{\bm x} \cdot {\bm z}_{{\bm x}+\hat\mu}|^2 =
   {1\over 2} \sigma_{\bm x} \sigma_{{\bm x}+\hat\mu} + {1\over2}.
\end{equation}
We obtain therefore an Ising Hamiltonian. 

For $N=3$, configurations should be labeled by the unit-length vector 
${\bm \tau} = {\bm s}_1 \times {\bm s}_2$ that encodes both the chirality of 
the two vectors and the plane in which they lie. 
The gauge-invariant tensor $T^{ab}_{L,\bm x}$ 
is related with ${\bm \tau}$ by $T^{ab}_{L,\bm x} = \frac{1}{2}
\sum_c \epsilon^{abc} \tau^c_{\bm x}$. Substituting in Eq.~(\ref{HTT}) 
we obtain 
\begin{equation}
  |\bar{\bm z}_{\bm x} \cdot {\bm z}_{{\bm x}+\hat\mu}|^2 =
   {1\over 2} {\bm \tau}_{\bm x}\cdot {\bm \tau}_{\bm x+\hat\mu} + 
   {1\over 4} ({\bm \tau}_{\bm x}\cdot {\bm \tau}_{\bm x+\hat\mu})^2 + 
   {1\over 4}.
\end{equation}
We thus obtain a ferromagnetic Heisenberg model.

These results are confirmed by a standard LGW analysis. In the
critical limit, the Hamiltonian with hopping term (\ref{HTT}) becomes
equivalent to the LGW model for an antisymmetric $N\times N$ field
$\Psi^{ab}({\bm x})$, which represents the coarse-grained average of
$T_{L,\bm x}^{ab}$.  The corresponding LGW Lagrangian reads
\begin{eqnarray}
  {\cal L} &=&  {\rm Tr} \, \partial_\mu \Psi^t \partial_\mu \Psi
  + r \,{\rm Tr} \,\Psi^t\Psi\nonumber\\
&+&   \,u \, ({\rm Tr}
\,\Psi^t\Psi)^2 + v\, {\rm Tr}\, (\Psi^t\Psi)^2,
\label{hlgt} 
\end{eqnarray}
where we have written the quadratic terms in terms of the transpose 
$\Psi^t$ (since $\Psi$ is antisymmetric
$\Psi^t=-\Psi$) to show explicitly their positivity. Note that the cubic
term is absent because ${\rm Tr} \Psi^n = 0$ for any odd $n$.

We can easily recover the results obtained in the large-$v$ limit for
$N=2$ and 3, For $N = 2$ we can write $\Psi^{ab}$ in terms of a single
real scalar field $\phi$, i.e., $\Psi^{ab} = \epsilon^{ab}\phi$. The
two quartic terms are equivalent, and we obtain the LGW model for a
real scalar field that is associated with the Ising universality
class.  For $N=3$ we can write $\Psi^{ab}(x)$ in terms of a single
three-component vector as $\Psi^{ab} = \epsilon^{abc}\phi^c$ where
$\epsilon^{abc}$ is the completely antisymmetric tensor. Again, the
quartic terms are equivalent and we obtain the O(3) vector LGW
Hamiltonian.  Thus, continuous transitions should belong to the O(3)
vector universality class.  No simplifications occur for $N\ge 4$. To
determine the critical behavior one should therefore study the RG flow
of the model (\ref{hlgt}) in the space of the quartic couplings $u$
and $v$, for which there are no known results in the literature.

Assuming, as usual, that the LGW analysis is valid for any $v$,
the previous results indicate that, for $N=2$, an Ising behavior 
should also occur for finite positive values of $v$, down to the point $v=0$, 
where the symmetry enlarges to
$SU(2)$ and the model undergoes a Heisenberg transition.  The
finite-$v$ behavior for $N=2$ can also be understood by considering a
variant model in which the $SO(N)$-symmetric potential
$V_O$  is added to the standard
$CP^{N-1}$ Hamiltonian $H_{CP}$ defined in  Eq.~(\ref{standardCPN}).  For
this purpose we parametrize the fields in terms of a real 
three-component unit vector ${\bm t}_{\bm x}$ defined by 
\begin{equation}
   t_{k,\bm x} = \bar{\bm z}_{\bm x} \sigma_k {\bm z}_{\bm x},
   \label{tPauli}
\end{equation}
where $\sigma^k$ ($k=1,2,3$) represent here the Pauli matrices.  It
is then easy to verify that
\begin{eqnarray}
 &&  H_{CP} + v \sum_{\bm x} \,|{\bm z}_{\bm x}\cdot {\bm z}_{\bm x}|^2 
\nonumber \\ 
 && =
  - {1\over 2} N J \sum_{{\bm x}\mu} ({\bm t}_{\bm x}\cdot {\bm t}_{{\bm x}
    +\hat\mu} + 1) + v \sum_{\bm x} (1 - t_{2,\bm x}^2).
\end{eqnarray}
Thus, the Hamiltonian is equivalent to a
Heisenberg model with a symmetry breaking term $- v t_{2,\bm
  x}^2$. For positive $v$, only the second component of ${\bm t}$
becomes critical, so that the transition is in the Ising universality
class. On the other hand, for negative $v$ the system magnetizes in
the (1,3) plane, undergoing a $XY$ transition, as already discussed.
Transitions for $v=0$ correspond to O(3)-symmetric multicritical
points, belonging to the Heisenberg universality class (accurate
estimates of the critical exponents and other universal features can
be found in
Refs.~\cite{Hasenbusch-20,Chester-etal-20-o3,HV-11,CHPRV-02}), where
the $XY$ transition line for $v<0$ and the Ising transition line for
$v>0$ meet.

To understand the finite-$v$ behavior for $N=3$, we recall that the
lattice $CP^2$ model undergoes a first-order transition, see
e.g., Ref.~\cite{PV-19-CP}, so that it is natural to expect first-order
transitions also for small positive values of $v$. As a consequence,
we predict the existence of a tricritical value $v^*$, such that the
transition is in the Heisenberg universality class for $v > v^*$ and
of first order for $v < v^*$.

We do not further discuss the more complicated cases with $N\ge 4$.
Further work is clearly necessary to clarify their critical  behavior 
for positive values of $v$.

\subsubsection{The CH transition line}

As CM transitions, also CH transitions are characterized by the
spontaneous breaking of the $SO(N)$ symmetry. As discussed in
Sec.~\ref{sec.2A}, the symmetry breaking pattern depends on the sign
of $v$ and thus different Higgs phases are obtained for $v > 0$ and $v
< 0$.  However, at variance with the CM transitions, along the CH line
the gauge-field modes are expected to play an active role, since the
CH line separates a Coulomb phase with long-range gauge modes from a
Higgs phase with massive gauge excitations.  Therefore, as in the
ULAHM~\cite{BPV-21-ncAH,BPV-22}, the critical behavior along the CH
line is expected to be described by the OAHFT (\ref{OAHFT}), which can
be obtained by taking the formal continuum limit of the OLAHM.

The analysis of the RG flow of the OAHFT reported in
Sec.~\ref{oahftsec} shows that a stable charged fixed point only
exists for $N$ larger than a critical value $N^*(d=3)$, unless new
universality classes emerge in three dimensions that are unrelated
with the RG flow close to four dimensions, 
as it occurs for the one-component AH
models~\cite{BFLLW-96,NRR-03,BPV-23c}.  In analogy with what occurs in
the ULAHM~\cite{BPV-21-ncAH,BPV-22}, we expect $N^*(d=3)$ to be of
order 10.  This fixed point is located in the region $v > 0$ and is
therefore different from the charged fixed point that controls the
critical behavior of the ULAHM, which belongs to the line $v=0$.

The field theory results allow us to predict the nature of the CH
transitions.  For $N<N^*(d=3)$, all transitions along the CH line are
expected to be of first order, as no stable charged fixed point
exists. For $N>N^*(d=3)$ the nature of the transition depends on the
sign of $v$. Since the line $v=0$ is a separatrix of the RG flow, and
the charged fixed point lies in the region $v>0$, the charged fixed
point is unreachable for systems with negative $v$. Thus, CH
transitions that separate a Coulomb disordered phase from a Higgs
phase with residual ${\mathbb Z}_2\otimes O(N-1)$ 
invariance---this is the symmetry
breaking pattern characterizing systems with $v<0$---are expected to
be of first order. Continuous transitions can only be observed in
systems with $v > 0$, provided that the system is in the attraction
domain of the stable fixed point.

For $\kappa=\infty$ the gauge variables freeze, thus we can fix all
gauge variables to the trivial value $\lambda_{{\bm x},\mu}=1$,
obtaining an $O(2)\otimes O(N)$ model with Hamiltonian
\begin{eqnarray}
  H_{O,\kappa\to\infty}
  &=& - 2NJ
\sum_{{\bm x},\mu} {\rm Re}\,(\bar{\bm z}_{\bm
  x} \cdot {\bm z}_{{\bm x}+\hat\mu}) \nonumber\\
&+& v
  \sum_{\bm x} \,|{\bm z}_{\bm x}\cdot {\bm z}_{\bm x}|^2.
 \label{LAHHkinfty}
\end{eqnarray}
As already mentioned in Sec.~\ref{oahftsec}, the corresponding LGW
theory has 3D stable fixed points, and therefore continuous
transitions are possible in the lattice model
(\ref{LAHHkinfty}). However, as already discussed in
Sec.~\ref{oahftsec}, any stable fixed point of the purely scalar
theory is unstable with respect to gauge fluctuations. Therefore, for
finite $\kappa$, one can never observe the same asymptotic critical
behavior as for $\kappa = \infty$. However, the $\kappa=\infty$
criticality may give rise to substantial pre-asymptotic crossover
effects for finite large values of $\kappa$.

\section{Conclusions}
\label{conclu}

We have investigated some generalizations of the standard
multicomponent AH models with $SU(N)$ symmetry, such as the AH field
theory defined in Eq.~(\ref{SAHFT}) and the lattice AH model defined
in Eq.~(\ref{SLAH}).  By adding an appropriate scalar potential, we
obtain models with a reduced symmetry, which, therefore, may undergo
transitions belonging to different universality classes and have Higgs
phases with different symmetries.  In this paper we focus on AH models
with $SO(N)$ invariance, the field theory with
Lagrangian~(\ref{OAHFT}) and the lattice AH model with
Hamiltonian~(\ref{OLAH}). We determine the possible symmetry-breaking
patterns and the symmetry of the Higgs phases, finding that they
depend on the sign of the Hamiltonian parameter $v$.  Thus, for $v >
0$ and $v < 0$ we observe different ordered phases characterized by
the condensation of different gauge-invariant order parameters. In
particular, there are different Higgs phases.

The analysis of the RG flow of the OAHFT, using the perturbative and
large-$N$ computations of Refs.~\cite{Hikami-80,MU-90}, indicates that
the quartic scalar term that breaks the $SU(N)$ symmetry is a relevant
perturbation of the $SU(N)$-symmetric fixed point. Therefore, in the
absence of exact $SU(N)$ symmetry, under RG transformations the system
flows away from the $SU(N)$ fixed point, possibly toward a stable
fixed point--- in this case one may observe an $SO(N)$-symmetric
critical behavior---or toward infinity---in this case, a first-order
transition would occur.  The analysis of the RG flow close to four
dimensions and in the large-$N$ limit~\cite{Hikami-80} shows that an
$SO(N)$-symmetric stable fixed point exists for $N>N_O^*(d)$ and that
it lies in the region $v>0$. We expect this fixed point to exist also
in three dimensions for $N$ sufficiently large, i.e., for
$N>N_O^*(d=3)$ [by analogy with the $SU(N)$-symmetric case, we guess
  that $N_O^*(d=3)$ is of order ten].  It would control the behavior
of lattice systems at transitions where the gauge degrees of freedom
are critical. We should note, however, that we cannot exclude the
existence of 3D stable fixed points that are not related with the
fixed points identified by the $\varepsilon$ expansion close to four
dimensions, as it happens in the one-component 3D
ULAHM~\cite{BFLLW-96,NRR-03,BPV-23c}, and in LGW theories that are
effective models of the $O(2)\otimes O(N)$ scalar models that are
obtained by taking the $\kappa\to\infty$ limit of the multicomponent
OLAHM~\cite{DPV-04,CPPV-04}.

Concerning the 3D lattice models, we argue that some features of the
phase diagram of $SO(N)$-symmetric lattice AH models are independent
of the Hamiltonian parameter $v$ and similar to those of the
$SU(N)$-symmetric models.  In all cases, the qualitative phase diagram
is the one reported in Fig.~\ref{phadiaLAH}, with three different
phases---the Coulomb, the molecular, and the Higgs phase---which
differ in the properties of the gauge correlations, in the confinement
or deconfinement of the charged excitations, and in the residual
symmetry of the ordered (molecular and Higgs) phases.

As far as the nature of the phase transitions, we argue that MH
transitions always belong to the $IXY$ universality class, as in
$SU(N)$ invariant models.  This is due to the fact that MH transitions
are topological transitions only driven be gauge modes. Scalar fields
play no role at the transition.  On the other hand, the nature of the
CM and CH transitions depends on $v$ and different behaviors are
observed for positive and negative values of $v$. This is due to the
fact that the breaking of the $SO(N)$ symmetry differs for $v>0$ and
$v<0$ [the residual symmetry is $O(N-2)$ and ${\mathbb Z}_2\otimes
  O(N-1)$ in the two cases, respectively], with the condensation of
different order parameters, such as $R_{L,{\bm x}}^{ab}$ and
$T_{L,{\bm x}}^{ab}$ defined in Eqs.~(\ref{wqoper}) and
(\ref{wqopera}).

The behavior along the CH line is essentially determined by the
presence or absence of the field-theoretical fixed point.  For $N <
N^*(d=3)$ we expect only first-order transitions along the CH line.
For $N > N^*(d=3)$, if the Higgs phase is symmetric under ${\mathbb
  Z}_2\otimes O(N-1)$ transformations, which is the residual symmetry
for $v<0$, the CH line is again a line of first-order
transitions. Indeed, if the RG flow starts in a point $v < 0$, it
necessarily runs toward infinity.  Instead, if the Higgs phase is
$O(N-2)$-symmetric, i.e., $v$ is positive, CH transitions may be
continuous, in the universality class associated with the field-theory
fixed point.

Also the behavior along the CM line depends on $v$. For negative $v$,
transitions are driven by the condensation of $R_{\bm x}^{ab}$,
defined in Eq.~(\ref{wqoper}). Gauge modes play no role, so that one
can perform a simple LGW analysis to determine the nature of the phase
transitions.  It predicts first-order transitions for any $N\ge
3$. For $N=2$ continuous transitions in the $XY$ universality class
are possible. For positive $v$, we argue that the critical behavior is
effectively described by the $O(N)$-symmetric LGW theory for an
antisymmetric rank-two tensor, which is the coarse-grained analogue of
$T_{\bm x}^{ab}$, see Eq.~(\ref{hlgt}). This allows us to predict
that, for $N=2$, CM transitions should belong to the Ising
universality class for all positive values of $v$. For $N=3$, instead,
we expect the existence of a tricritical value $v^*>0$, such that the
transition is of first order for $v < v^*$ and in the Heisenberg
universality class for $v > v^*$. The existence of a tricritical value
is due to the first-order nature of the CM transitions in $SU(3)$
invariant AH models, i.e., for $v = 0$.  The behavior for $N\ge 4$ is
not known, since no results for the RG flow of the corresponding LGW
theory are available.

Of course, numerical checks of the above predictions would be useful
and welcome.  However, we believe that the theoretical arguments
reported in this paper, that also rely on the known behavior of
$SU(N)$ invariant AH models (which have been carefully studied
numerically, see, e.g., Refs.~\cite{BPV-21-ncAH,BPV-22,BPV-23c}), are
sound and can be easily extended to AH models with more general scalar
interactions. We also stress that the predictions reported above
should not only apply to the OLAHM, but also to more general models,
in which the unit-length constraint for the lattice variable ${\bm
  z}_{\bm x}$ is relaxed.

We remark that one may also consider more general quartic
scalar potential. For instance, one may consider the AH field theory
with quartic potential~\cite{MU-90}
\begin{eqnarray}
V_{{\mathbb P}}({\bm\phi}) =
  u  \,(\bar{\bm\phi} \cdot {\bm\phi})^2 
+ v \,|{\bm\phi}\cdot {\bm\phi}|^2 
+ w \,\sum_{a=1}^N (\bar\phi^a\phi^a)^2,
\label{Vphip}
\end{eqnarray}
which in only invariant under the permutation group ${\mathbb P}_N$.
A one-loop analysis of the RG flow close to four dimensions was
  reported in Ref.~\cite{MU-90}, showing that a stable fixed point
  appears only for very large values of $N$, more precisely for $N\ge
  5494$.  Note that for $N=2$ the quartic potential $V_{{\mathbb
    P}}({\bm\phi})$ is the most general one preserving the $U(1)$
gauge invariance and the uniqueness of the quadratic $\bar{\bm
  \phi}\cdot {\bm\phi}$ term.  For $N>2$ there are other quartic terms
satisfying these conditions, such as $\sum_{a=1}^N \bar\phi^a \phi^a
\bar\phi^{a+1} \phi^{a+1}$ (we identify $\phi^{N+1} = \phi^1$), which
leaves a residual ${\mathbb Z}_N$ symmetry only. Analogous terms can
be added to the lattice AH models.

Lattice AH counterparts with residual ${\mathbb P}_N$ global
  symmetry can be simply obtained by adding a term $w \sum_{\bm x}
  \sum_{a=1}^N (\bar{z}^az^a)^2$ to the $SO(N)$-symmetric Hamiltonian
  (\ref{OLAH}).  We expect the phase diagram of these lattice AH
models (keeping the quartic parameters fixed) to be qualitatively the
same as that of $SU(N)$- and $SO(N)$-symmetric theories, with three
phases and three transition lines, as sketched in
Fig.~\ref{phadiaLAH}. MH transitions are always expected to belong to
the $IXY$ universality class. Instead, the nature of the transitions,
and the universality classes of the continuous ones along the CM and
CH lines, are expected to change.  In particular, in the $N=2$
  lattice AH model with quartic potentials analogous to $V_{{\mathbb
      P}}({\bm\phi})$, the residual symmetry is ${\mathbb
    P}_2={\mathbb Z}_2$. Thus, we expect CM transitions to be Ising
  transitions.  This behavior can be easily confirmed by rewriting the
  ${\mathbb Z}_2$-symmetric scalar potential in terms of the variable
  (\ref{tPauli}), obtaining $V_P({\bm z}) = v (1 - t_2^2) + {1\over2}
  w (1 + t_3^2)$.  Thus, for generic values of $v$ and $w$ the system
  undergoes Ising transitions. On the planes $v=0$, $w=0$, and
  $2v+w=0$ one can observe both Ising and $XY$ transitions, depending
  on the symmetry of the low-temperature phase. The nature of the CH
  transitions is less clear, but we believe these transitions to be of
  first order, since no stable fixed points are found in the
  corresponding AH field theory, at least close to four
  dimensions~\cite{MU-90}.  Although we believe that the phase diagram
  and critical behaviors of these extended lattice AH theories are
  worth being investigated, we have not pursued this study further.

We finally stress that the understanding of the possible extensions of
the AH gauge theories, allowing for more general scalar potentials,
may be useful to get a more thorough understanding of the possible
phases and critical behaviors that can be observed in critical
phenomena characterized by an emerging Abelian gauge field.

\end{document}